\documentclass[epjCONF, onecolumn]{svjour}
\usepackage{graphics}
\usepackage[varg]{txfonts} 
\usepackage[latin1]{inputenc}
\session-title{Assembling the Puzzle of the Milky Way}
\begin{document}
\title{The Milky Way and other spiral galaxies}
\author{F. Hammer\inst{1}\fnmsep\thanks{\email{francois.hammer@obspm.fr}} \and M. Puech\inst{1} \and H. Flores\inst{1} \and Y. B. Yang\inst{1} \and J. L. Wang\inst{1}  \and S. Fouquet\inst{1} }
\institute{GEPI, Observatoire de Paris, CNRS, 5 Place Jules Janssen, 92195 Meudon, France}
\abstract{
Cosmologists have often considered the Milky Way as a typical spiral galaxy, and its properties have considerably influenced the current scheme of galaxy formation.
Here we compare the general properties of the Milky Way disk and halo with those of galaxies selected from the SDSS. Assuming the recent measurements of its circular velocity results in the Milky Way being offset by $\sim$2$\sigma$ from the fundamental scaling relations. On the basis of their location in the ($M_{K}$, $R_{d}$, $V_{flat}$) volume, the
fraction of SDSS spirals like the MilkyWay is only 1.2\% in sharp contrast with M31, which appears to be quite typical. 
Comparison of the Milky Way with M31 and with other spirals is also discussed to investigate
 whether or not there is a fundamental discrepancy between their mass assembly histories. Possibly the Milky Way is one of the very few local galaxies that could be a direct descendant of very distant, z=2-3 galaxies, thanks to its quiescent history since thick disk formation. } 
\maketitle
\section{Introduction}
\label{intro}
The Milky Way is one of the 72\% of massive\footnote{By massive galaxies, we arbitrarily consider those with stellar masses larger than $10^{10}M_{\odot}$, the Milky Way being five times more massive than this value.} galaxies that are disk dominated. How large disks formed in massive spirals? The question is still not fully answered. Disks are
supported by their angular momentum that may be acquired by early interactions in the framework of the tidal
torque (TT) theory  \cite{Peebles76,White}. In this theory, galactic disks are then assumed to evolve without subsequent major mergers, in a
secular way, as did the Milky Way from its early and gradual disk formation. Is the Milky Way an archetype of spiral galaxies? The answer is of crucial importance for galaxy formation theory because the TT theory faces with at least two major problems. First, galaxy simulations demonstrate that such disks
can be easily destroyed by collisions \cite{Toth92}, and such collisions might be too frequent to let
disks survive. Second, the disks produced by simulations are too small or have a too small angular
momentum when compared to the observed ones, the so-called "spin catastrophe".

\section{How the Milky Way compares to other local (SDSS) spirals?}
\label{sec:1}
 It is now well
established that the Milky Way experienced very few minor mergers and no major merger during the past 10-11 Gyrs
\cite{Wyse,Gilmore}. The old stellar content of the thick disk let possible a merger origin at such an early epoch, which is still a matter of debate. The Milky Way is
presently absorbing the Sagittarius dwarf, though it is a very tiny event given that the Sagittarius mass is
less than 1\% of the Milky Way mass \cite{Helmi}. 

\begin{figure}
\resizebox{0.98\columnwidth}{!}{
\includegraphics{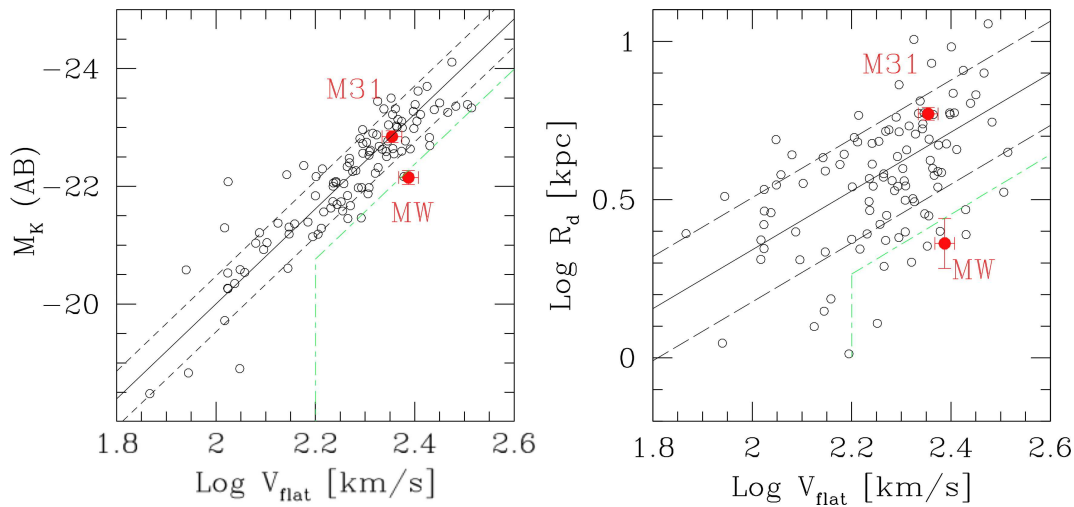}}
\caption{Reproduction of the Figure 5 of \cite{Hammer07} with the new measurements for the circular velocity of the Milky Way. One-$\sigma$ uncertainty of both relations
is shown as dashed lines. Long- and short-dashed lines show how we select
Milky Way like galaxies, which are discrepant in both $L_{K}$ and disk scalelength.}
\label{fig:1}       
\end{figure}

How the fundamental parameters
of the Milky Way disk compare to those of other galaxies? The main observational difficulty is coming from the fact that we are lying into the Milky Way,
and such a comparison is not an easy task. By chance, very detailed models of the light distribution of the Galaxy were
necessary to remove at best its signal to recover the CMB emission. Hipparcos also provided very useful data to
model in detail the Galaxy. Using these data, \cite{Hammer07} have shown that the Galactic disk scalelength, $R_{d}$= 2.3 $\pm$ 0.6 kpc,
is quite small especially when compared to that of M31 ($R_{d}$= 5.8 $\pm$0.6 kpc). The whole emission of both the Milky
Way and M31 in K-band have been well recovered by Cobe \& Spitzer, and provides $ M_{K}(AB)$=-22.15 and -22.84, for the Milky
Way and M31, respectively. The difference between the two values indicates that the stellar mass of M31 is twice
that of the Milky Way, after accounting for their respective stellar mass to K-band luminosity ratios. Even if the
Milky Way is approximately twice gas rich than M31, the baryonic mass ratio is still close to 2, because the gas content is
rather marginal in both galaxies. 

On the basis of a very detailed
study of the local scaling relations (mass-velocity or Tully Fisher, radius-velocity) for local spirals, \cite{Hammer07}
showed that M31 is quite a typical spiral, while the Milky Way is surprisingly exceptional, being offset by 1$\sigma$ in
both relations. The new measurement by \cite{Reid09} with data being reanalysed by \cite{Bovy09} provides a Milky Way velocity of 244 km/s instead of 220 km/s as adopted by \cite{Hammer07}.  Fig. 1 shows how the position of the Milky Way in the K-band Tully Fisher and in the $R_{d}$-$V_{flat}$ relationships, together with M31 and SDSS galaxies from a complete sample of \cite{Pizagno07}.  If correct, the new velocity for the Milky Way would be in excess of that of M31, and would place the Milky Way at $\sim$ 2$\sigma$ for both relations. We have searched for SDSS galaxies with comparable masses that would share the same location than the Milky Way in the ($M_{K}$, $R_{d}$, $V_{flat}$) volume. Only one galaxy (SDSS235607.82+003258.1) among the 79 SDSS galaxies with $Log(V_{flat})\ge$ 2.2 shows a position similar to that of the Milky Way.  Thus only 1.2$\pm$1.2\%\footnote{Error has been calculated as in \cite{Hammer07}, and assuming that the \cite{Pizagno07} sample is representative of SDSS galaxies; the distribution of galaxies around the mean of both relations shows a similar scatter than found by other studies.} of SDSS galaxies share the location of the Milky Way in that volume, i.e., significantly smaller than the value (7\%) found by \cite{Hammer07}. Examination of Fig. 1 shows that this is mainly due to the offset of the Milky Way in the well-defined Tully Fisher relation (only two SDSS galaxies show a similar offset) and not to the scale-length estimate of the Milky Way, as it has been previously argued by \cite{van der Kruit11}. 

 In addition, \cite{Mouhcine06} showed that Milky Way stars in the inner halo have much bluer colour than corresponding stars in haloes of other spirals, including M31, which implies a deficiency in their [Fe/H] abundances by almost 1 dex. Helmi (2011, private communication) argued that having an external view of the Milky Way would change this result because of the prominence of the Sagittarius Stream. On the other hand the Sagittarius Stream represents a small fraction of the halo stellar mass \cite{Hammer07}, and its [Fe/H]= -1.2 abundance \cite{Sesar11} is still offset by -0.5 dex when compared to M31 and other spiral haloes. Thus, as firstly guessed by Allan Sandage and verified by \cite{Flynn06} and \cite{Hammer07}, it appears that the Milky Way is almost certainly an exceptional spiral.

\section{How the past history of the Milky Way compares to that of other spirals?}
Let us first consider M31. Quoting Sidney van den Bergh\cite{vandenBergh05} in his introduction of the book "The Local Group as an Astrophysical Laboratory": ``Both the high metallicity of the M31 halo, and the $r^{1/4}$ luminosity profile of the Andromeda galaxy, suggest that this object might have formed from the early merger and subsequent violent relaxation, of two relatively massive metal-rich ancestral objects.'' In fact the considerable amount of streams in the M31 haunted halo could be the result of a major merger \cite{Hammer10} instead of a considerable number of minor mergers. This alternative scenario provides a robust explanation of the Giant Stream (GS) discovered by \cite{Ibata01}: observed properties of GS stars are consistent with tidal tail stars that are captured by the gravitational potential of a galaxy after a major merger.  In fact GS stars have ages older than 5.5 Gyr \cite{brown07}, which is difficult to reconcile with a recent collision, such as expected for a minor merger \cite{Font08}. The stellar age constraint has let \cite{Hammer10} to reproduce the M31 substructures (disk, bulge \& thick disk) as well as the GS after a 3:1 gas-rich merger for which the interaction and fusion may have occurred 8.75$\pm$0.35 and 5.5 $\pm$0.5 Gyr ago, respectively.

M31 being a quite typical spiral and possibly a major merger relics, one may wonder what is the general past history of most spirals, and how it differs from that of the Milky Way.   Progenitors of present-day giant spirals are similar to galaxies having emitted their light $\sim$ 6 Gyr ago, according to the Cosmological Principle. Since the Canada France Redshift Survey \cite{Hammer95,Hammer97}, observations of distant galaxies up to z$\sim$ 1 can provide data with depth and resolution comparable to what is currently obtained for local galaxies. Six billion years ago, the Hubble sequence was very different from the present-day one \cite{Delgado09}. While the E/S0 number density shows no evolution, more than half of the spiral progenitors show peculiar morphologies. Furthermore, \cite{Neichel08} demonstrated that in addition to their peculiar morphologies, these galaxies show anomalous velocity fields at large scales (7 kpc) from their extended ionised gas, i.e. not consistent with rotation.  This indicates a common process perturbing the ionised gas and stars in half of the spiral progenitors. This cannot be caused by outflows since in most cases there are no velocity shift between emission and absorption lines  \cite{Hammer09}. Most anomalous galaxies reveal peculiar large-scale gas motions that cannot be caused by minor mergers or by secular evolution (e.g. bars), both mechanisms resulting in too small and/or too spatially localised kinematic perturbations \cite{Puech07,Puech11}. Internal fragmentation should have a limited impact at these redshifts ($z_{mean}$=0.65): less than 20\% of the IMAGES sample \cite{Yang08} show clumpy morphologies according to \cite{Puech10} while associated cold gas accretion tends to vanish in massive halos at z$<$1, with $<$1.5 $M_{\odot}$/yr at z$\sim$ 0.6 \cite{Keres09}.

Besides this, \cite{Hammer09} succeeded to reproduce the morpho-kinematics of the anomalous galaxies based on a grid of simple major merger models based
on \cite{Barnes02}, providing convincing matches in about two-thirds of the cases. This suggests that a third of z=0.4-0.75 spiral galaxies are
or have been potentially involved in a major merger. Why so many major mergers? In fact the morpho-kinematic observational technique used in IMAGES is found to be sensitive to all merger phases, from pairs to post-merger relaxation.  \cite{Puech11} has compared the merger rate associated with these different phases, and found a perfect match with predictions by state-of-the-art $\Lambda$CDM semi-empirical models \cite{Hopkins10} with no particular fine-tuning.  Thus, both theory and observations predict an important impact of major mergers in the progenitors of present-day spiral galaxies.

\section{Conclusion: what can we learn on and from the Milky Way formation?}

The specific angular momentum of the Milky Way is half that of spirals with similar velocities.  Could the quiescent past history of the Milky Way explain its lack of angular momentum? A significant part of the observed angular momentum of spiral galaxies may come from the orbital angular momentum generated by major mergers, which may solve the spin catastrophe \cite{Maller02}. However the Milky Way has still a too large angular momentum when compared to expectations from the TT theory. \cite{Hammer07} conjectured that  a significant part of its disk mass had been acquired through gas accretion, explaining its observed lack of angular momentum. This does not exclude a very ancient and gas-rich merger, a possibility supported by the distribution of orbital eccentricities of thick disk stars \cite{Dierickx09} . It can be also tested whether the bulge has a classical component \cite{Babusiaux10}, because only primordial collapse or merger are known to produce such a component. The reverse is not necessarily true: some gas-rich mergers may also produce bulges with low Sersic indices \cite{Wang11}.

There are now many simulations of disk formation via gas-rich mergers, including in the cosmological context \cite{Brook11}. They naturally produce an important thick disk component  that is made of material re-accreted by the newly  re-formed galaxy \cite{Brook04,Hammer10}. Constrained cosmological simulations of the Local Group are also progressing rapidly towards promising predictions \cite{Forero11}. The Milky Way having escaped such an event for at least 10 billion years, it could be an almost direct descendant of galaxies with a few $10^{10}M_{\odot}$ at z$\sim$ 2-3.

\end{document}